# Polygonal silica toroidal microcavity for controlled optical coupling

Takumi Kato, Wataru Yoshiki, and Takasumi Tanabe

Department of Electronics and Electrical Engineering, Faculty of Science and

Technology,

Keio University

3-14-1, Hiyoshi, Kohoku-ku, Yokohama, 223-8522, Japan

Tel/Fax: +81-45-566-1730/1529

takasumi@elec.keio.ac.jp

Abstract:

We fabricated polygonal silica toroidal microcavities to achieve stable mechanical coupling with an evanescent coupler such as a tapered fiber.  The polygonal cavity was fabricated by using a combination of isotropic etching, anisotropic etching and laser reflow.  It offers both high and low coupling efficiencies with the cavity mode even when the coupler is in contact with the cavity, which offers the possibility of taking the device outside the laboratory. A numerical simulation showed that an octagonal silica toroidal microcavity had an optical quality factor of $8.8 \times 10^6$.

In recent years, optical microcavities such as silica toroidal microcavities [1 2], silica microspheres [3], and photonic crystals [4], have been extensively studied [5] because they can yield a strong interaction between light and matter.  These cavities offer high quality factors ($Q$) and small mode volumes $V$, which enable low-power signal processing, cavity quantum electro-dynamics, and sensing.  Of the various types of microcavities, whispering gallery mode (WGM) cavities exhibit the highest $Q$ ($Q > 10^8$) [1]. Because the light confinement is achieved by total internal reflection at the side surface of WGM microcavities, technologies that reduce the surface roughness are particularly important if we are to obtain an ultrahigh $Q$. For this purpose, a laser reflow technique is often used to smooth the sidewalls. This approach removes the surface roughness by using the surface tension that occurs during the reflow process.

Of the various kinds of silica WGM microcavities, the ultrahigh-$Q$ silica toroidal microcavity is particularly attractive for use in applications such as optical frequency comb generation [6] and biomolecule sensing [7], because it can be fabricated on a chip. Although various demonstrations have been reported, all of the above applications have occurred in laboratories, and it remains a challenge to employ them in the field. One technological difficulty is to achieve the stable and accurate sub-µm control of the position of an evanescent coupler (i.e. tapered fiber [8, 9]), which is needed if we are to input/output the light to/from the cavity in a controlled way. If we can touch the coupler with the cavity sidewall, the alignment is easier but this usually results in overcoupling with the cavity mode. So, in this work, we developed a method for fabricating a polygonal (specifically octagonal) silica toroidal microcavity and studied its optical characteristics to enable us to achieve high and low optical coupling even when the tapered fiber is in contact with the cavity. To the best of our knowledge, this is the first demonstration of the manipulation of the shape of a high-$Q$ toroidal microcavity to control the external coupling. Changing the microcavity shape is a simple and straightforward way of adjusting the optical coupling between a microcavity and couplers, as has been demonstrated with silicon (Si) microrings and microdiscs [10, 11]. However, it should be noted that only round silica toroidal microcavities have been considered because of the $CO_2$ laser reflow that is required during the fabrication process. First, we describe a method for fabricating polygonal toroidal microcavities, and then we describe our analysis of the polygonal cavity using a two-dimensional finite-difference time-domain (FDTD) simulation.

As shown in Fig. 1, an important step when fabricating a toroidal microcavity is the laser reflow, where we expose a $CO_2$ laser beam from the top of the disk cavity. The laser liquefies the $SiO_2$ and after the diffusion of the heat to the Si post, the cavity rim solidifies, and the surface becomes smooth as a result of surface tension. This process is crucial for reducing the scattering loss and achieving an ultrahigh $Q$. Since surface tension makes the structure round, thus far only circular toroidal microcavities have been fabricated. The key to achieving a unique shape is to control the shape of the underlying etched Si layer.

Figure 1 shows the steps involved in fabricating a toroidal microcavity. First, a $SiO_2$ disk cavity is fabricated on a silicon wafer by thermal dioxidization, photolithography, and $SiO_2$ etching. The dioxide layer of our device is 3 μm thick. Next, the silicon sacrificial layer is undercut to form a Si post. Here we carried out a combination of isotropic and anisotropic etching to obtain a polygonal silicon post. And finally the $SiO_2$ is exposed with a $CO_2$ laser. Since the silicon layer has higher thermal conductivity than $SiO_2$, it functions as a heatsink during the laser reflow [1]. As a result, the shape of the silicon post is transferred to the $SiO_2$ when we employ a $CO_2$ laser beam with uniform intensity. This is the key to obtaining a smooth polygonal toroidal cavity.

The anisotropic wet etching is performed using KOH, which allows us to etch <100> and <110> crystal faces quickly. So, when we use a Si wafer with a <100> surface, the Si post will have eight flat faces, and consequently form an octagon. To make a silicon post with the desired shape and diameter, we performed the KOH etching followed by a $XeF_2$ isotropic etching. (We also investigated isotropic wet etching with a mixture of hydrofluoric, nitric, and acetic acids (HNA), which also worked well for our purpose.) Isotropic etching is needed before anisotropic etching to obtain the sufficiently small silicon post diameter needed to form a toroidal cavity. Indeed, we obtained an almost perfect octagonal post as shown in Fig. 2(a). However, without isotropic etching but with increased KOH etching time we obtained a different structure as shown in Fig. 2(b). It is difficult to obtain an octagonal post with a large undercut without isotropic etching. With our structure, we performed the isotropic etching before the KOH etching to reduce the post diameter to 80 μm from 100 μm (i.e. the undercut of one side is 10 μm). Then we used a 48% KOH etchant to form the octagonal shape. The anisotropic etching time was 3 hours. Although KOH etching cannot make silicon perfectly octagonal, the combined use of ethylene diamine and pyrocatechol (EDP) or tetramethylammonium hydroxide (TMAH) and changing the mask pattern from circular to elliptic should make it possible to obtain a perfect octagon and other polygonal forms.

Figure 2(c) shows a fabricated polygonal toroidal cavity. The disk cavity (Fig. 2(a) right) with the polygonal silicon post was exposed with a 100-ms $CO_2$ laser pulse operating at a power of 14 W. The diameter of the focus spot of the laser beam was about 120 μm. This step reflow the $SiO_2$ and makes the side wall smooth, which should support a sufficiently low scattering loss. At the same time, the polygonal shape of the silicon post is transferred to the $SiO_2$ layer because Si works as a heatsink.

Since the shape of the cavity does not depend on the shape of the $CO_2$ laser beam but on the shape of the Si post, it should be possible to obtain a different shape by controlling the shape of the Si post.

To confirm that our structure will support controlled coupling efficiencies with the waveguides while exhibiting a sufficiently high $Q$, we performed a numerical analysis based on two-dimensional FDTD. The cavity dimensions are shown in Fig. 3(a). We modeled a curved octagonal shape because the edges are smoothed by the laser reflow. First we calculated a toroidal cavity with a standard circular shape and a radius of 20 μm as a reference, with which we obtained a $Q$ of more than $1\times10^7$ where the value is limited by the computation time. Next, we investigated the octagonal toroidal cavity. Figure 3(b) shows the mode spectrum, in which we observe clear cavity resonance with a free-spectral range of 5.2 nm. We investigated one of the high $Q$ modes and obtained a $Q$ of $8.8\times10^6$. The profile for this mode is shown in Fig. 3(c). The characteristic of this mode is slightly different from that obtained from a standard round cavity. A close view of the optical mode at the edge of the cavity (shown in the left panel of Fig. 3(c)) reveals that the light propagates close to the surface at the corner, but it propagates slightly inward at the side. This small difference results in a large change in the coupling strength when the cavity is placed with the coupler. Controlling the coupling is particularly important for practical applications. In addition it suggests that this cavity is more robust against surface contamination because light propagates more inside the cavity than in a round cavity.

So, we investigated the coupling of this polygonal WGM cavity when we placed a tapered fiber, which is commonly used for coupling, close to the cavity sidewall. The loaded $Q$ ($Q_{load}$) of a cavity is given by

$$Q_{load}^{-1} = Q_{unload}^{-1} + Q_{couple}^{-1} \qquad (1)$$

where $Q_{unload}$ and $Q_{couple}$ are the unloaded $Q$ and the coupling $Q$, respectively. $Q_{load}$ is a $Q$ with input/output waveguides and $Q_{unload}$ is that without a waveguide. $Q_{couple}$ is obtained from $Q_{load}$, $Q_{unload}$, calculated from FDTD, and Eq. (1).

We investigated $\kappa$ for different $d$ values and different coupling positions, where $\kappa$ is the coupling coefficient given as $\kappa = \sqrt{\omega/Q_{couple}}$, and $d$ is the gap between the cavity surface and the fiber. Calculated mode profiles for two different coupling positions are shown in Fig. 4(a). Here we have two options regarding the position of the tapered fiber; namely, at a straight part or at a corner.

Fig. 4(b) shows the coupling coefficient $\kappa$ with respect to $d$ for three different cases. In a circular toroidal microcavity, there is no way to control $\kappa$ other than to change $d$. On the other hand, the coupling strengths differ for an octagonal cavity when the waveguide is placed parallel to the straight part of the cavity (parallel coupling) or at the corner (corner coupling). It is difficult to achieve precise distance control with the tapered fiber due to the mechanical instability that occurs easily outside the laboratory. With this in mind, the easiest way to achieve stable coupling is to bring the fiber into contact with the cavity surface. Although we can only have one value for a round cavity, we can obtain two different $\kappa$ values for a polygonal cavity at $d = 0$. Namely, polygonal cavities have an advantage when it comes to controlling the coupling strength because we can choose the coupling position.

We have a lower coupling coefficient for parallel coupling than for corner coupling at $d = 0$, ($\kappa = 65$ ms$^{-1/2}$ (parallel coupling) and 200 ms$^{-1/2}$ (corner coupling)), which is consistent with the explanation in Fig. 3(c). Since the cavity charging and discharging time $\tau_c$ is given by $\tau_c = \kappa^{-2}$, the $\tau_c$ value is 0.24 μs for parallel coupling and 25 ns for corner coupling. This suggests that we can charge and discharge the cavity 10 times faster or slower simply by changing the coupling configuration without making the system mechanically unstable. It also changes the coupling efficiency, since the transmittance at the resonance is given by $T = (Q_{load}/Q_{unload})^2$, where $Q_{load}$ is dependent on $\kappa$.

Figure 4(b) also shows another advantage of a polygonal cavity compared with a round cavity. The coupling coefficient does not decrease monotonously as $d$ increases in a circular cavity mode, whereas it does in a polygonal cavity mode. Since the evanescent field decays exponentially, $\kappa$ should decrease exponentially with respect to $d$. The unexpected behavior seen in Fig. 4(b) indicates that the refractive index perturbation becomes so large that the mode profile of the circular cavity is modified when we place the fiber too close to the cavity. On the other hand, even though the mode profile shown in the right panel in Fig. 4(a) exhibits noticeable modification for the corner coupling, Fig. 4(b) suggests that the perturbation is still small, and the $\kappa(d)$ dependence remains exponential. It suggests that this system is robust against perturbation from the outside environment.

In summary, we have developed a novel method for fabricating a smooth polygonal silica toroidal microcavity by combining the isotropic and anisotropic etching of a sacrificial silicon layer. We also showed that WGM-like modes are obtained in an octagonal cavity, and that these modes propagate in the cavity with a path that is separated more from the surface at a straight part of the cavity. This enables us to have a smaller $\kappa$ at the straight part, and a higher $\kappa$ at the corner. This in turn leads to better controllability of the coupling strength with the waveguides, in particular when the evanescent coupler is in contact with the cavity, which contributes to the robustness of the system. Our approach requires only simple packaging and enables us to take this device out of the laboratory, which will significantly extend the application of toroidal microcavities.

Part of this work was supported financially by the Strategic Information and Communications R&D Promotion Programme (SCOPE), the Toray Science and Technology Grant, and Keio University's Program for the Advancement of Next Generation Research Projects.

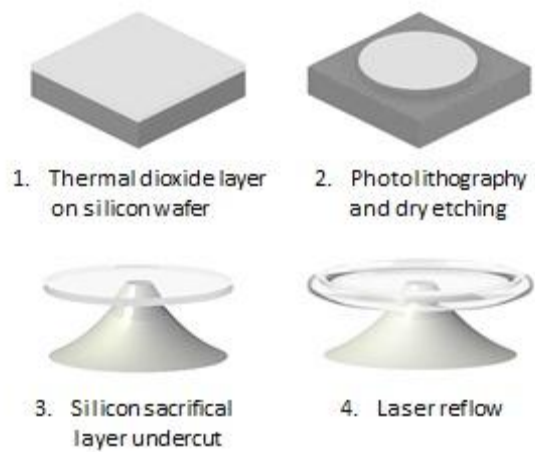

Fig. 1: (color online) Process for fabricating a toroidal microcavity.

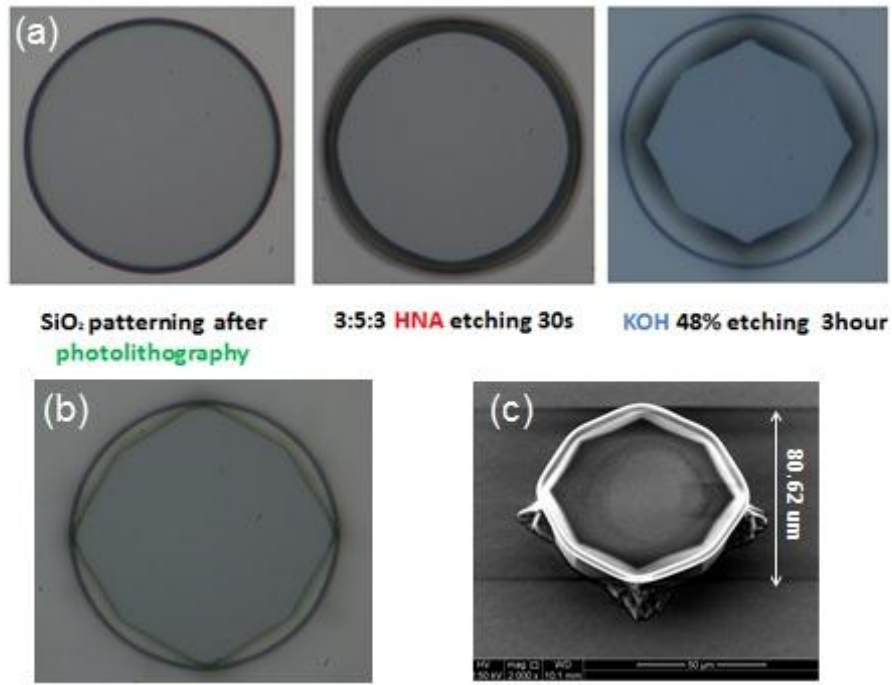

Fig. 2: (color online) (a) Optical microscopic images of a disk cavity to form an octagonal silicon post. (b) Optical microscope image of a disk cavity when KOH etching is performed for 4 hours 30 minutes and without isotropic etching. (c) Scanning electron microscope image of a fabricated octagonal toroidal microcavity after the laser reflow.

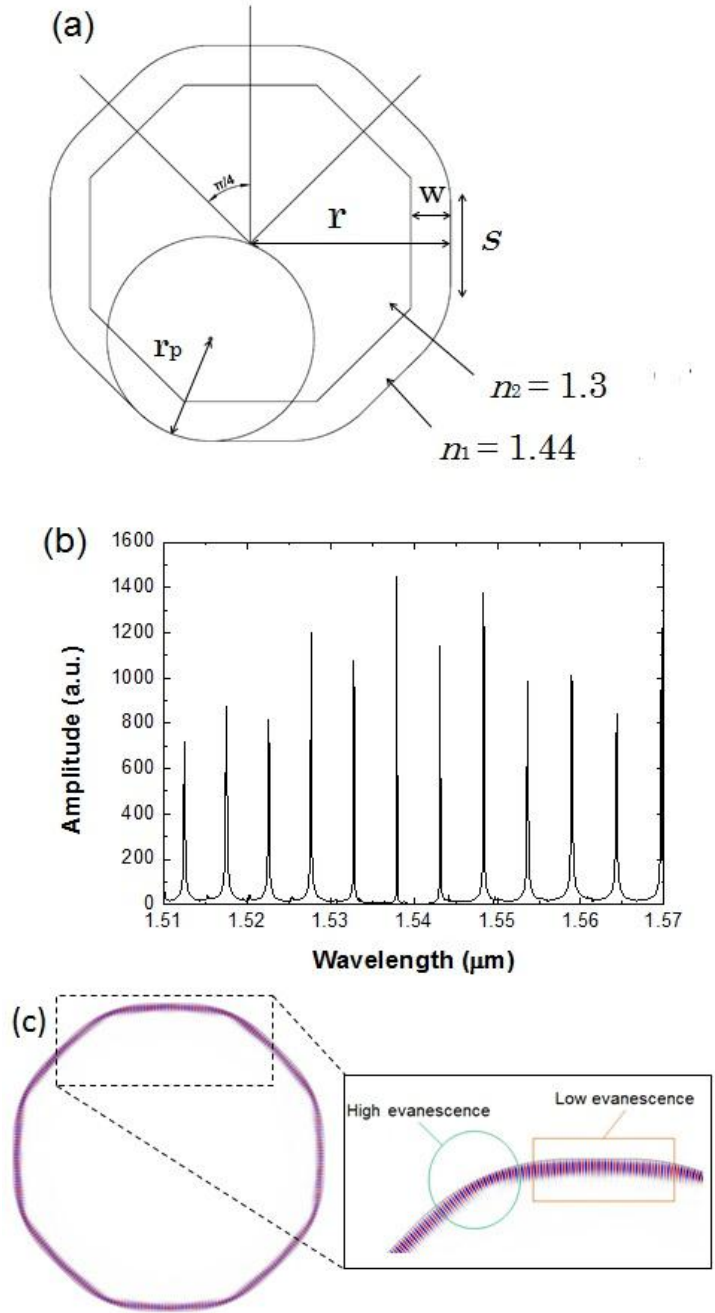

Fig. 3: (color online) (a) Structure and dimensions of the polygonal toroidal microcavity used for the calculation. The cavity radius $r$ is 50 μm, and the cavity rim width $w$ is 10 μm. The vertex curvature and the side length of the polygon are $r_p$ = 38.1 μm and $s$ = 10 μm, respectively. The effective refractive index of the SiO$_2$ at the rim is $n_1$ = 1.44 and that of the remaining part is $n_2$ = 1.3 assuming a 1-μm thick slab. (b) The resonance spectrum of the octagonal silica toroidal microcavity shown in (a).  (c) $H_z$-field profile of an octagonal toroidal microcavity.  The inset shows a close-up of the mode profile at the corner and the side of the octagon.

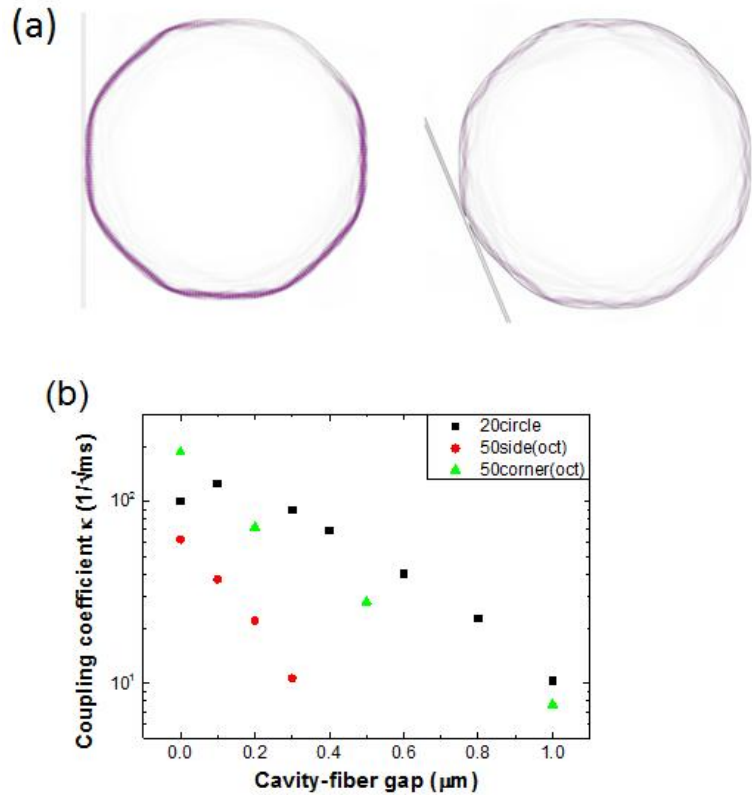

Fig. 4: (color online) (a) Calculated mode profiles for a hexagonal toroidal microcavity (shown in Fig. 3(a)) with a tapered fiber differently touched to the surface.  The diameter of the tapered fiber is 1 μm.  (b) Coupling coefficient $\kappa$ with respect to $d$ for a circular cavity with $r = 20$ μm and for a hexagonal cavity with two different coupling configurations.  Square, round, and triangular dots show $\kappa$ for a circular cavity, parallel coupling ((a) left) and corner coupling ((a) right), respectively.